\documentclass{article}
\usepackage{arxiv}

\usepackage[utf8]{inputenc} 
\usepackage[T1]{fontenc}    
\usepackage{hyperref}       
\usepackage{url}            
\usepackage{booktabs}       
\usepackage{amsfonts}       
\usepackage{nicefrac}       
\usepackage{microtype}      
\usepackage{lipsum}		
\usepackage{graphicx}
\usepackage{doi}

\usepackage{cite}
\usepackage{amsmath,amssymb,amsfonts}
\usepackage{algorithmic}
\usepackage{graphicx}
\usepackage{textcomp}

\usepackage{placeins}
\usepackage{lineno,hyperref}
\modulolinenumbers[5]
\usepackage[linesnumbered,ruled, vlined]{algorithm2e}
\usepackage{url}
\usepackage{multicol}
\usepackage{multirow}
\usepackage{caption}
\usepackage{subcaption}

\usepackage{scalerel}
\usepackage{tikz}

\title{Deep Learning-based Framework for Automatic Cranial Defect Reconstruction and Implant Modeling}

\author{ \href{https://orcid.org/0000-0002-8076-6246}{\includegraphics[scale=0.06]{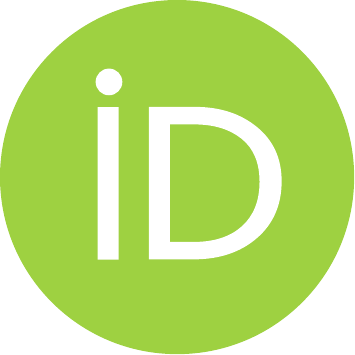}\hspace{1mm}Marek Wodzinski}\thanks{Corresponding author.} \\
    AGH University of Science and Technology \\
    Department of Measurement and Electronics \\
	Krakow, Poland \\
	University of Applied Sciences Western Switzerland \\ Information Systems Institute \\
	Sierre, Switzerland \\
	MedApp S.A., Krakow, Poland \\
	\texttt{wodzinski@agh.edu.pl} \\
	\And
	\href{https://orcid.org/0000-0003-2363-7912}{\includegraphics[scale=0.06]{orcid.pdf}\hspace{1mm}Mateusz Daniol} \\
	AGH University of Science and Technology \\
    Department of Measurement and Electronics \\
	Krakow, Poland \\
    MedApp S.A., Krakow, Poland \\
	\And
	\href{https://orcid.org/0000-0001-9462-8269}{\includegraphics[scale=0.06]{orcid.pdf}\hspace{1mm}Miroslaw Socha} \\
    AGH University of Science and Technology \\
    Department of Measurement and Electronics \\
	Krakow, Poland \\
	\And
	\href{https://orcid.org/0000-0002-2193-7690}{\includegraphics[scale=0.06]{orcid.pdf}\hspace{1mm}Daria Hemmerling} \\
    AGH University of Science and Technology \\
    Department of Measurement and Electronics \\
	Krakow, Poland \\
	\And
	\href{https://orcid.org/0000-0002-6764-3815}{\includegraphics[scale=0.06]{orcid.pdf}\hspace{1mm}Maciej Stanuch} \\
	AGH University of Science and Technology \\
    Department of Measurement and Electronics \\
	Krakow, Poland \\
    MedApp S.A., Krakow, Poland \\
    \And
	\href{https://orcid.org/0000-0003-2299-458X}{\includegraphics[scale=0.06]{orcid.pdf}\hspace{1mm}Andrzej Skalski} \\
	AGH University of Science and Technology \\
    Department of Measurement and Electronics \\
	Krakow, Poland \\
    MedApp S.A., Krakow, Poland \\
}

\renewcommand{\shorttitle}{\textit{Preprint}}

\hypersetup{
pdftitle={Deep Learning-based Framework for Automatic Cranial Defect Reconstruction and Implant Modeling},
pdfsubject={},
pdfauthor={Marek Wodzinski},
pdfkeywords={Augmented Reality, AutoImplant, Cranial Implant Design, Deep Learning, Image Segmentation, Implant Modeling, Shape Completion, Skull Reconstruction, Personalized Medicine},
}

\begin{document}
\maketitle

\begin{abstract}
\textit{Background and Objective} The goal of this work is to propose a robust, fast, and fully automatic method for personalized cranial defect reconstruction and implant modeling. 

\textit{Methods} We propose a two-step deep learning-based method using a modified U-Net architecture to perform the defect reconstruction, and a dedicated iterative procedure to improve the implant geometry, followed by automatic generation of models ready for 3-D printing. We propose a cross-case augmentation based on imperfect image registration combining cases from different datasets. We perform ablation studies regarding different augmentation strategies and compare them to other state-of-the-art methods.

\textit{Results} We evaluate the method on three datasets introduced during the AutoImplant 2021 challenge, organized jointly with the MICCAI conference. We perform the quantitative evaluation using the Dice and boundary Dice coefficients, and the Hausdorff distance. The average Dice coefficient, boundary Dice coefficient, and the 95th percentile of Hausdorff distance are 0.91, 0.94, and 1.53 mm respectively. We perform an additional qualitative evaluation by 3-D printing and visualization in mixed reality to confirm the implant's usefulness. 

\textit{Conclusion} We propose a complete pipeline that enables one to create the cranial implant model ready for 3-D printing. The described method is a greatly extended version of the method that scored 1st place in all AutoImplant 2021 challenge tasks. We freely release the source code, that together with the open datasets, makes the results fully reproducible. The automatic reconstruction of cranial defects may enable manufacturing personalized implants in a significantly shorter time, possibly allowing one to perform the 3-D printing process directly during a given intervention. Moreover, we show the usability of the defect reconstruction in mixed reality that may further reduce the surgery time.
\end{abstract}


\section{Introduction}
\label{sec:introduction}

The process of modeling cranial implants is important for neurosurgery. The implants are used to repair defects induced by craniectomy or other sources of cranial damage. It is necessary to automate the process of implants design and production, since nowadays the process takes days to weeks and induces the necessity of a follow-up intervention~\cite{autoimplant_1,design_1,design_2,design_3,design_4,li_3}. When designing the shape of the implant, it is important to adjust its surface shape and thickness to the location of the defect, as well as to adjust the edge trim, which will enable the formation of a bone-implant connection.

The process of cranial implant modeling can be improved and automated by deep learning (DL)~\cite{autoimplant_1,li_3}. The process may be treated as a segmentation (shape completion) task where the segmented (completed) volume is the cranial defect~\cite{morais_1}. In the DL-based approach, the computational load is transferred to the training phase and the inference process is usually fast enough to be performed in real-time. It enables one to reconstruct the defect during surgery and to refine it to create an implantable model. The implant model can be then 3-D printed directly during the intervention, thanks to the recent advances in medical 3-D printing~\cite{print_1,print_2,print_3}.

This work focuses primarily on the automatic reconstruction of cranial defects and implant modeling, and presents applications of the calculated models in 3-D printing and augmented reality (AR)~\cite{ar}.

\subsection{Related Work}

There are several important contributions to the automatic design of cranial implants in the literature. In this work, we focus on the most recent advances. For the detailed discussion about older contributions, we refer to the summary of the 1st edition of the AutoImplant challenge~\cite{autoimplant_1}.

The 1st edition of the AutoImplant challenge inspired researchers to propose several unique solutions. The approaches were inspired mostly by DL~\cite{autoimplant_1}. The challenge organizers provided a baseline method~\cite{li_1}. The contributions were primarily based on an encoder-decoder or U-Net-like architecture~\cite{39,40,ellis_1,42,43,45,46}, however, one method applied 2-D GANs~\cite{pimentel_1} and presented that statistical shape models (SSM) may reduce the risk of overfitting and improve the generalizability. The best performing teams applied both the skull preprocessing and the defect augmentation~\cite{autoimplant_1,ellis_1,42,43,40,39}. Three contributions used shape priors with convolutional neural network, UNet architecture and variational auto-encoder applied with encoder-decoder network~\cite{39,40,45}. We hypothesize that the use of shape prior is beneficial for small datasets. However, with strongly augmented larger datasets the network should be able to learn it by themselves. The winner of the 1st edition presented that extending the training set by image registration may improve the quantitative results~\cite{ellis_1}. They concluded that augmenting the training set is crucial for the shape completion problem and we follow this idea in our research.

Several important contributions were introduced during the AutoImplant 2021 challenge~\cite{autoimplant_2}. The challenge consisted of three separate tasks: (i) Task 1 related to the reconstruction of prealigned, variously shaped defects, (ii) Task 2 considering the generalizability into real, clinical cases, and (iii) Task 3 related to the improvements in skull shape completion (using the same dataset as during the 1st edition of the challenge). One of the contributions decided to train two 3D U-Net-like networks separately on Task 1/3~\cite{mahdi_1}. They used the model trained on Task 1 in Task 2, to show its generalizability, and used second-step filtering to increase the reconstruction of fine details. Another interesting contribution was based on implant prediction using the slice-by-slice approach~\cite{yang_1}. The researchers used a recurrent neural network (RNN) to use the \textit{a priori} knowledge about the continuity of the adjacent slices. The work presented in~\cite{yu_1} addressed the challenge differently. The author argued that generalizability is crucial in cranial implant design, and proposed a method based on the principal component analysis (PCA), using only a subset of Task 3 data for training. Moreover, the authors used a registration-based approach to create a common image domain. Considering that a small training set was used, the method achieved superior results on Task 2, connected with the generalizability into real cranial defects. The authors of~\cite{pathak_1} participated in Task 3 only and proposed a two-step procedure involving initial bounding box search and refinement in higher resolution. Both the steps applied the 3-D V-Net. Our preliminary work~\cite{wodzinski_1} outperformed the other methods. We have shown that appropriate data augmentation and linking are crucial to obtain reasonable results on different distributions. Two additional contributions came from the challenge organizers. The submissions attempted to solve the problem related to high GPU memory requirements of the DL-based contributions~\cite{li_2,kroviakov_1}. In~\cite{li_2}, the authors observed that majority of the input voxels are uninformative. They argued that an autoencoder may be used to learn a voxel rearrangement limiting the required memory. On the other hand, in~\cite{kroviakov_1}, authors decided to apply the spare convolutional neural networks~\cite{choy_1}. Using this approach, they excluded empty voxels from computations and decreased memory consumption, allowing one to use a higher resolution input, without the necessity of downsampling. 

There are also recent contributions not directly linked to AutoImplant challenge objectives~\cite{memon2021review}. In~\cite{kodym_1}, the authors proposed a DL-based solution to perform the skull segmentation from computed tomography (CT) scans. The method is based on 3-D U-Net, with results further refined by graph-cut. The method was used to create the Task 1 dataset, in contrast to Task 2 and Task 3 datasets, which were created using global thresholding. Even though no significant benefits were observed related to cranial implant design, the proposed method may be beneficial for other, more irregular, and complex structures. Another work elaborates on the neurosurgeons' criteria for cranial implant feasibility~\cite{ellis_2}. The authors proposed a scoring system, used to qualitatively evaluate the Task 2 submissions. They studied the correlation between different qualitative measures and scores from experts. They concluded that the automatically modeled implants require further manual refinement and that additional post-processing may improve the defects to meet given clinical requirements. Another work~\cite{rauschenbach_1} discussed the general insights into cranial implant design, motivation, manufacturing process, and currently used materials. The authors also presented an idea that the implant may be modeled and fabricated before the intervention, based on patient-specific pathological conditions.

\subsection{Contribution}

The referenced works suffer from the generalizability-related issues~\cite{autoimplant_2}. In this work, we present our contribution to the automatic cranial defect reconstruction and implant modeling using a combination of traditional and deep-learning approaches. We introduce an automatic, complete pipeline allowing one to create 3-D printing files from binary volumes representing the skull with the cranial defect. We evaluate our method on public datasets introduced for the 1st and 2nd editions of the AutoImplant challenge~\cite{autoimplant_1,autoimplant_2}. The presented method scored the first prize in all the challenge tasks. The key finding is that combining different datasets by \textit{imperfect} registration improves the training dataset diversity and outperforms other augmentation methods. Noteworthy, our method is trained purely on synthetic data, yet is still able to perform the skull reconstruction on real cranial defects. Moreover, we verify the 3-D printing capability, present a use case in augmented reality, and freely release the source code~\cite{source_code}.

\section{Methods}
\label{sec:methods}

\begin{figure*}[!htb]
	\centering
    \includegraphics[scale=0.65]{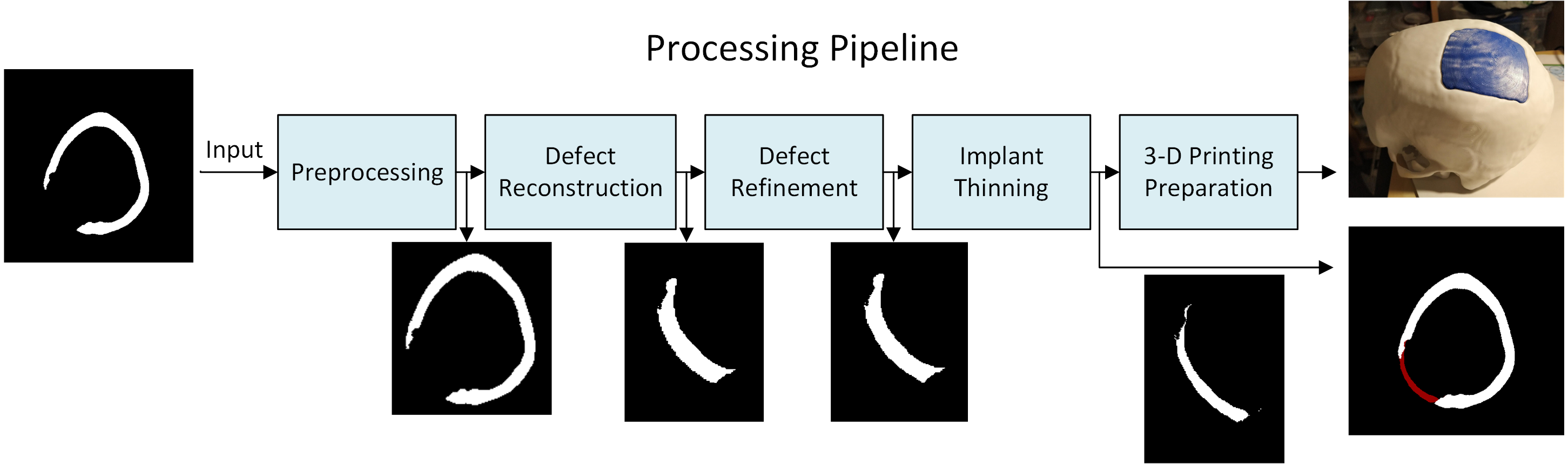}
    \caption{Visualization of the processing pipeline. Please note that the defect reconstructions are zoomed in for the presentation clarity.}   
    \label{fig:pipeline}
\end{figure*}

\subsection{Overview}

The proposed method is a complete pipeline to perform cranial defect reconstruction, implant modeling, and preparation for 3-D printing. The pipeline consists of several steps: (i) data loading and preprocessing, (ii) deep learning-based defect reconstruction, (iii) deep learning-based defect refinement, (iv) optional implant thinning, and (v) the preparation for 3-D printing. Additionally, the steps performed to link and augment the datasets are discussed separately since they have a significant influence on the reconstruction quality. The overview of the processing pipeline is shown in Figure~\ref{fig:pipeline}.

\subsection{Dataset}

\begin{figure}[!htb]
	\centering
    \includegraphics[scale=1.2]{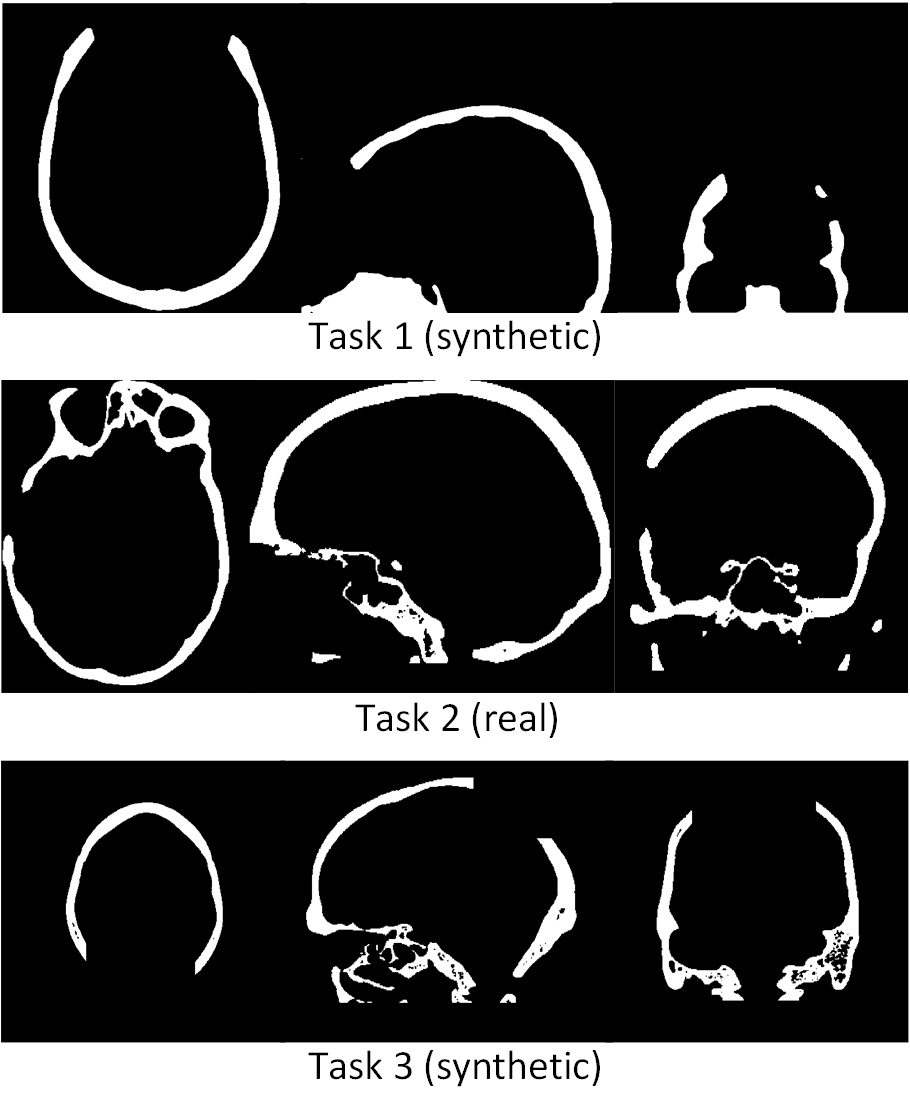}
    \caption{Exemplary cases from the datasets introduced for Task 1, 2 and 3, respectively. Please note differences induced by distinct skull segmentation techniques and the defect synthesis (Task 1, 3).}   
    \label{fig:datasets}
\end{figure}

The method is evaluated on the public datasets introduced during the AutoImplant 2021 challenge~\cite{autoimplant_2}. Three datasets are used: (i) SkullBreak (Task 1), real cranial defects (Task 2) and SkullFix (Task 3)~\cite{dataset}. The Task 1 and Task 3 datasets were adapted from a public head CT collection CQ500~\cite{cq_500} by the challenge organizers by segmenting the skulls and creating synthetic defects~\cite{dataset}. The comparison of cases from all the datasets is shown in Figure~\ref{fig:datasets}.

The Task 1 dataset is represented by 570 training cases created from 114 skulls and 100 testing cases created from 20 skulls. For each skull five defects are introduced: (i) bilateral, (ii) frontoorbital, (iii) parietotemporal, (iv) random, type 1, and (v) random, type 2. The goal of this task is to reconstruct defects with random shapes and positions, however, with already prealigned skulls. The ground-truths for the dataset are the synthetic defects.

The Task 2 dataset consists of 11 skulls with real cranial defects. The dataset may be considered as a benchmark for the generalizability into different distributions and the implant modeling. The dataset does not include any training cases. The ground-truths for the dataset are the manually created implants.

The Task 3 dataset is created from 100 skulls for training and 110 skulls for testing. Each skull is deteriorated with one synthetic defect. The test set is further divided into 100 skulls with defects similar to the training set and 10 skulls with defects from different distributions. The ground-truths for the dataset are the synthetic defects.

\subsection{Preprocessing and Dataset Linking}

The preprocessing starts with finding the boundaries of the defective skull. Then, the images are cropped with a predefined offset. The offset is required because the defect may extend beyond the bounding box of the defective skull. Then, the images are resampled to the same physical spacing (1 mm x 1 mm x 1 mm) and padded to the same, predefined shape (240 x 200 x 240). The steps are applied to cases from all datasets and allow one to use both the training sets together. The datasets are further offline augmented by cross-case image registration (IR) and variational autoencoder (VAE), discussed in sections~\ref{lab:ir} and \ref{lab:ig} respectively. The dataset linking and augmentation are crucial to improving the method performance, more important than the details of network architecture or the training hyperparameters.

\begin{figure*}[!htb]
	\centering
    \includegraphics[scale=0.63]{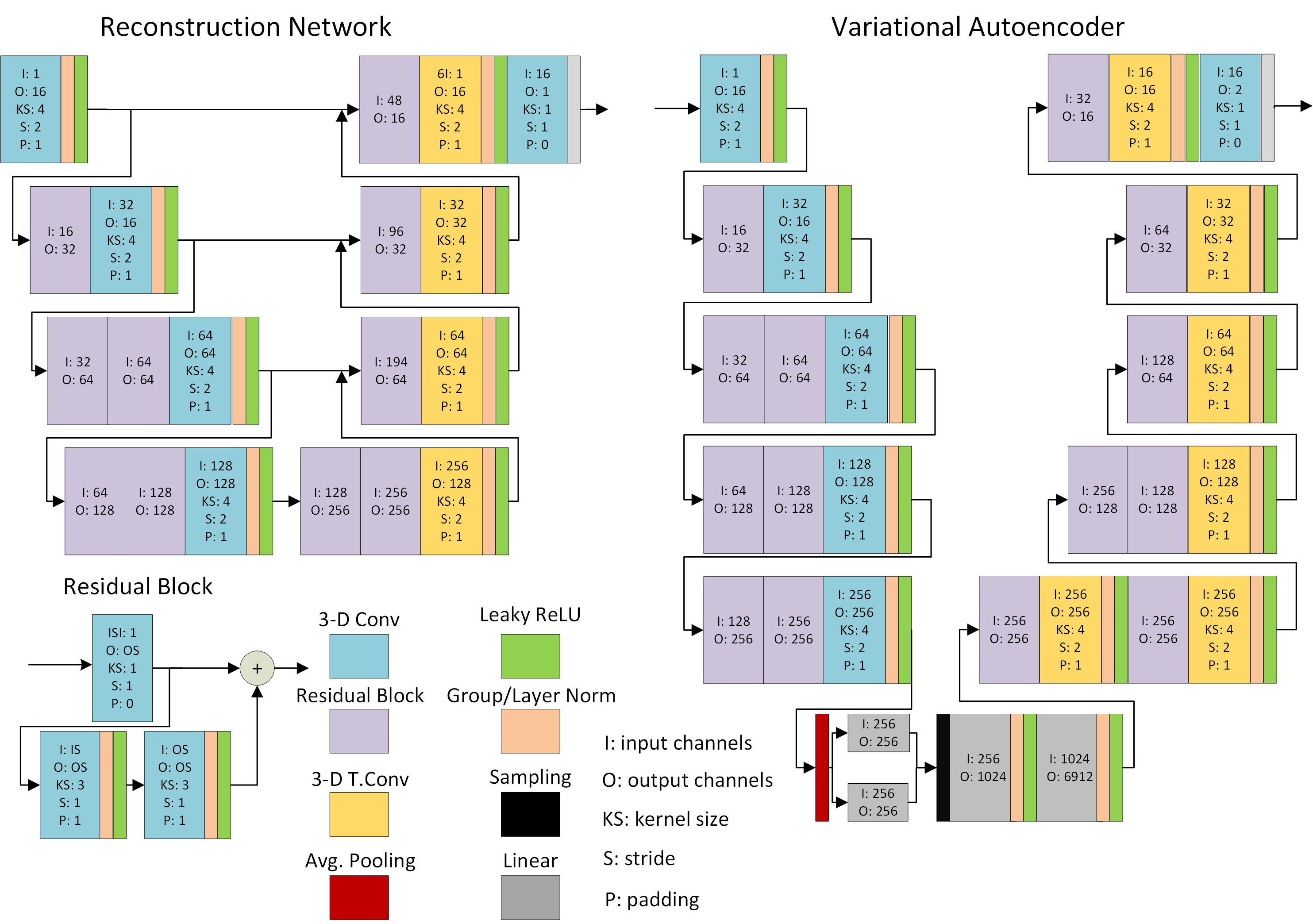}
    \caption{Visualizations of the architectures of the reconstruction/refinement network and the variational autoencoder.}   
    \label{fig:architecture}
\end{figure*}

\subsection{Defect Reconstruction and Refinement}

The defect reconstruction is performed by a U-Net-like network~\cite{unet} with residual blocks. The network takes as the input the preprocessed defective skull and outputs the calculated defect. The architecture is shown in Figure~\ref{fig:architecture}. We decided to calculate the defect instead of the complete skull due to slightly higher invariance to noise and smoother defect boundaries. The network is trained with the soft Dice loss as the cost function, defined as:
\begin{equation}
    DC_{loss}(A, B) = 1 - 2\frac{A \cap B}{A + B + \alpha},
\end{equation}
where A, B, $\alpha$ are the calculated defect, the ground-truth defect and the smoothing coefficient, respectively.

The defect refinement is an optional step performed after the reconstruction. The goal of the refinement is to smooth the calculated defect, make the upsampling errors negligible, and learn how to recover fine details that are unavailable at lower resolution. The refinement model uses the same architecture as the reconstruction, however, takes as the input the already calculated defect with additional preprocessing. First, the calculated defect is resampled and unpadded back to the original volume shape. Then, the bounding box for the reconsidered defect is calculated, cropped with a predefined offset (equal to 10 voxels in each dimension), and resampled to the same shape ($200^3$). During training, the same bounding box, padding, and resampling are used for the ground-truth defect. The defect refinement is trained the same way as the defect reconstruction. The intuitive effect of the defect refinement is shown in Figure~\ref{fig:refinement}.

\subsection{Augmentation by Image Registration}
\label{lab:ir}

The training sets are augmented by cross-case IR. All complete skulls from training sets are registered to each other. The registrations are performed using a multi-level, instance optimization-based approach~\cite{wodzinski_2}. First, an affine transformation is calculated and then followed by deformable IR. Since the registration is performed on binary images, the mean squared error (MSE) is used as the dissimilarity function. The diffusion regularization is used to regularize the nonrigid step. The cost function is defined as:
\begin{equation}
    C_{REG}(S, T, u) = MSE(S \circ u, T) + \theta Reg(u),
\end{equation}
where $S, T$ are the moving and fixed complete skulls, $u$ is the calculated displacement field, $\theta$ denotes the regularization coefficient, $Reg$ is the diffusive regularization, and $\circ$ denotes the warping operation. The displacement field $u$ is used to warp all images connected with the source image (the complete skull, the defective skull, and the implant) creating a new training case. The created cases are saved and added to the training set. The number of required registrations grows quadratically with the number of training cases.

Two training sets are created. One with registrations performed until convergence, with strong regularization and diffeomorphism enforcement by scaling and squaring~\cite{voxelmorph} (further denoted as smooth and invertible IR), and second performed with strongly relaxed regularization coefficient, without enforcing the deformation field invertibility, and for a predefined number of iterations (further denoted as imperfect IR). Importantly, the approach can be used to augment training sets for other segmentation and classification problems, as long as morphologies of the objects of interest are similar.

\subsection{Augmentation by Image Generation}
\label{lab:ig}

The training set created by smooth and invertible IR is used to train a VAE network to perform further augmentation by image generation. We use an encoder-decoder architecture with residual blocks, shown in Figure~\ref{fig:architecture}. The network is trained until convergence with the objective function defined as:
\begin{equation}
    C_{VAE}(\cdot) = DC_{loss}(I, G) + \beta KL(E, \mu, \sigma) - DC_{loss}(G_{S}, G_{I}),
\end{equation}
where $I, G$ are the input and the generated case respectively, $KL$ denotes the Kullback–Leibler (KL) divergence between the latent space distribution and the normal distribution, $\beta$ controls the influence of KL divergence, $E$ is embedding sampled from the latent space distribution, $\mu,\sigma$ are the parameters of the latent space distribution, and $G_{S}, G_{I}$ are the generated defective skull and the generated defect respectively.

\begin{figure}[!htb]
	\centering
    \includegraphics[scale=1.00]{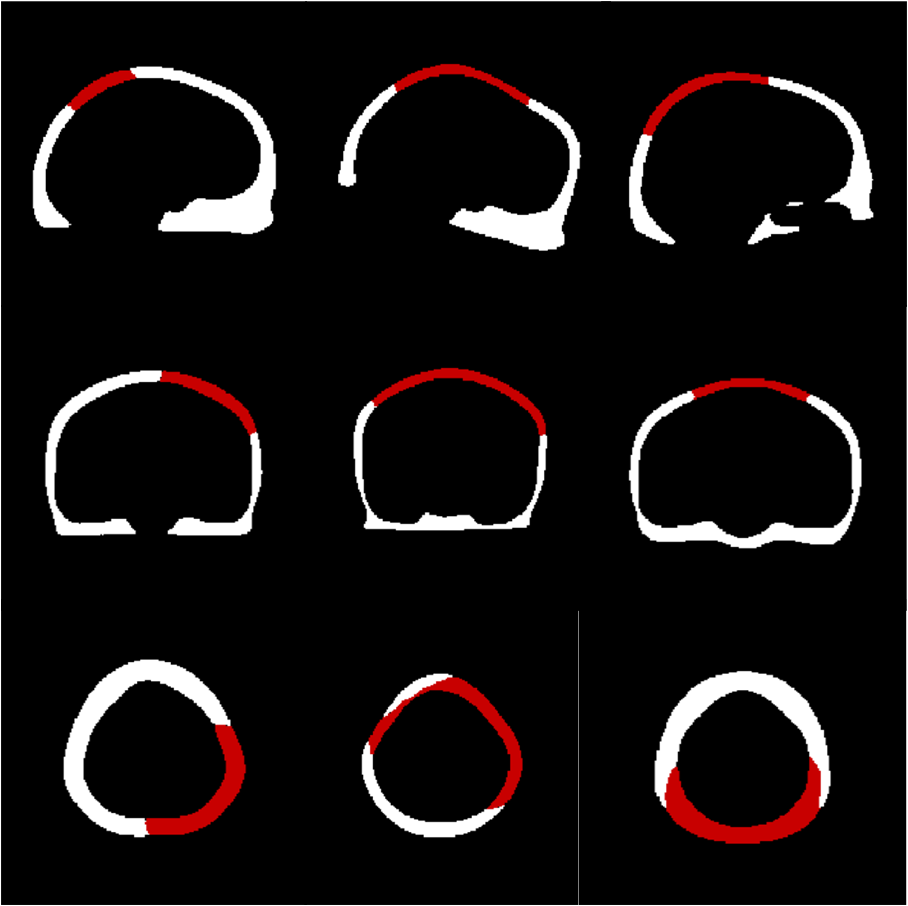}
    \caption{Exemplary training cases generated by the VAE.}  
    \label{fig:vae}
\end{figure}

We decided to use the VAE instead of other generative models because the desired similarity is well-defined and the generated images are binary. Therefore, the issues connected with blurring and incorrect fine details at image boundaries are not influential. The pretrained model is then used to create additional 100,000 training cases. The exemplary outputs of the VAE are shown in Figure~\ref{fig:vae}.

\begin{figure}[!htb]
	\centering
    \includegraphics[scale=0.70]{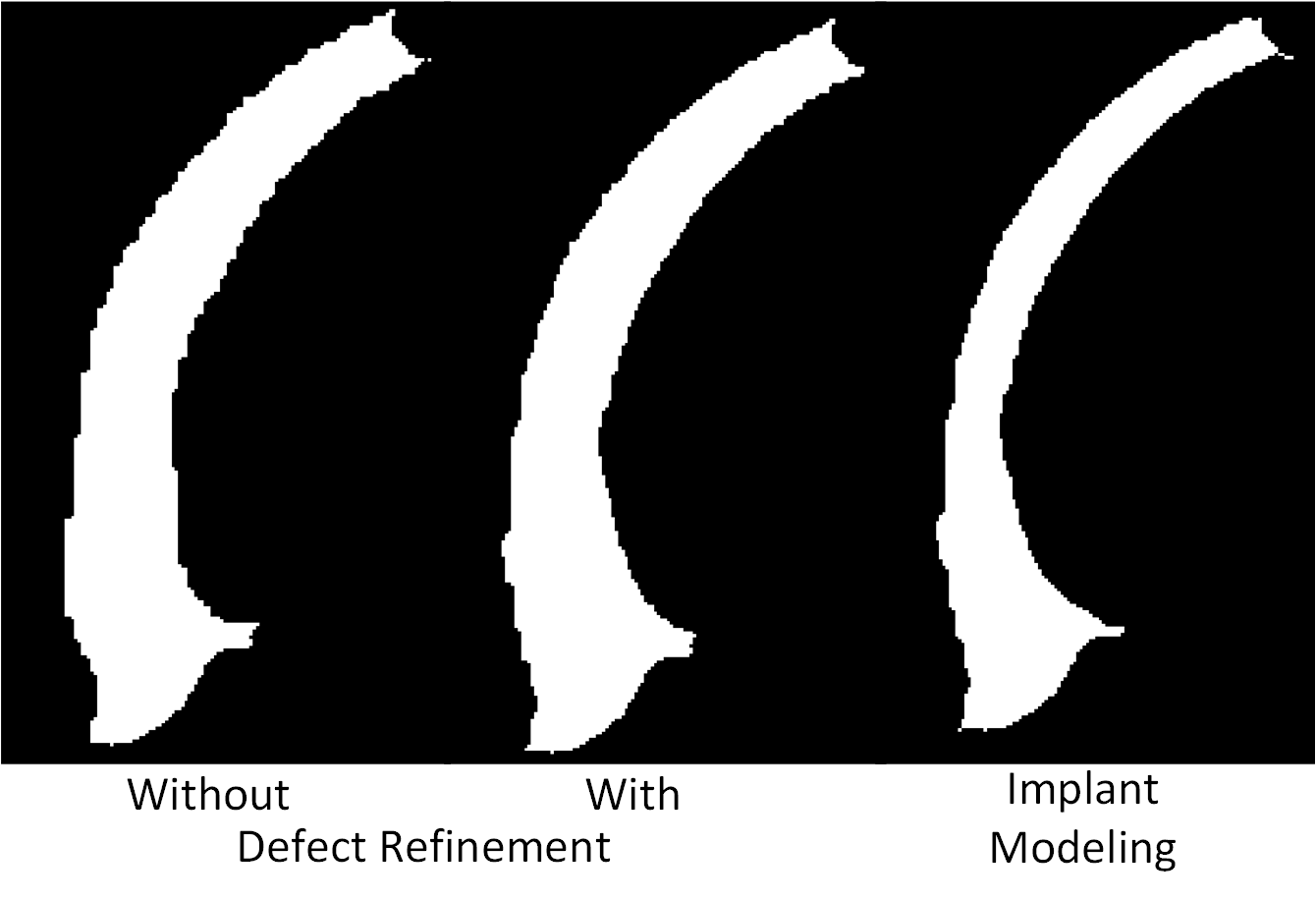}
    \caption{Visualization of the defect refinement and implant modeling. Please note that the refinement makes the reconstruction boundaries not only smoother but also captures fine details. The modeling makes the reconstruction thinner and implantable into the cranial cavity.}  
    \label{fig:refinement}
\end{figure}

\subsection{Postprocessing}

The postprocessing is performed to reverse the geometry change during the initial preprocessing and to decrease the influence of noisy reconstructions or interpolation artifacts. First, the given case is unpadded using parameters calculated during the preprocessing. Then, it is upsampled back to the original voxel size and unpadded again to the original shape. Then, we perform morphological postprocessing consisting of binary closing and connected component analysis. The binary closing together with exclusive disjunction is used to improve the implant boundaries. The connected component analysis is used to find the largest defect. This step is optional and can be tuned to the desired number of defects. We introduced it because we found that the ground-truth is always stated for only the largest defect, even if more are present.

\subsection{Implant Modeling}

The reconstructed defect usually cannot meet the clinical requirements related to its thickness and borders regularity~\cite{kodym_2,ellis_2}. Therefore, the reconstructed defects should be additionally postprocessed to create implantable implants. We propose a simple, iterative, and tuneable procedure to refine the defect.

The procedure starts from calculating the binary contour of the defect using exclusive or between the defect and the defect after the binary erosion. Then, we calculate two centroids, one of the defective skull, and the second of the calculated contour, followed by the calculation of a normalized vector between them. The defect is iteratively transformed along the vector with a predefined step. During each iteration, after the transformation, the logical conjunction between the original defect and the transformed defect is calculated, followed by median filtering and exclusive disjunction. Finally, the connected component analysis is performed to delete potentially introduced artifacts and to calculate the relative volume ratio between the transformed and the original defect. The process terminates after the desired volume ratio is achieved. The effect of the implant modeling is shown in Figure~\ref{fig:refinement}. In case of any errors, the implants may be further refined before manufacturing.

\subsection{3-D Printing}

We verify the designed defects and implants by 3-D printing. We perform the printing using Fused Deposition Modeling (FDM) technique and the general purpose Cartesian 3-D printer Prusa i3 MK3S+. For the test purposes, the objects are printed using PLA filament.

The models are postprocessed to make the printing process faster and smoother. The binary outputs are smoothed using Gaussian kernel, followed by isosurface creation using Flying Edges algorithm~\cite{flying_edges}. Finally, the mesh quality is improved by sinc-based filtering and connected component analysis to delete small, unconnected artifacts. 

The postprocessed output of implant modeling is converted to STL geometrical representation. The STL files are processed by PrusaSlicer software to create the 3-D printer G-Code instructions. Due to the technical limitations of the 3-D printer, the skulls are divided into two parts. This enables not only easier printing process, but also the evaluation of the cranial implant from the inner of the skull. As a result, the analysis of implant boundaries and implantability is easier.

\subsection{Experimental Setup}

The pipeline is implemented in Python using PyTorch library~\cite{pytorch}, extended by PyTorch Lightning framework~\cite{pylightning}. The postprocessing for 3-D printing is implemented using Python VTK library. The training cases are randomly split into training and validation sets with a 9:1 ratio. Only the training cases are used when combining the datasets or training the generative network. We perform several ablation studies concerning the training set structure:
\begin{itemize}
    \item The Task 1 training set only (T1)
    \item The Task 3 training set only (T3)
    \item The Task 1 and Task 3 training sets combined (Cmb)
    \item The Task 1 and Task 3 training sets combined by smooth and invertible IR (CReg) 
    \item The CReg with the defect refinement (CRegRef)
    \item The CReg further augmented by VAE (CRegVAE)
    \item The CRegVAE with the defect refinement (CRegVAERef)
    \item The Task 1 and Task 3 training sets combined by imperfect IR (CRegIm)
    \item An additional implant modeling (for Task 2 only) after the CRegIm (CImplant)
\end{itemize}
We also compare the presented pipeline to other state-of-the-art methods.

All the models are trained until convergence with batch size equal to 2, the number of cases per iteration is equal to 500, the initial learning rate being 0.003 and the decay rate varying from 0.95 to 0.99. Apart from the augmentation by IR or VAE, the training sets are online augmented by random affine transformations with scale, rotation, and translation ranging from 0.85 to 1.15, -15 to 15 degrees, and -10 to 10 voxels, respectively. Additional information about the source code, the experiments, and details on how to reproduce the results may be found in the repository~\cite{source_code}.

\section{Results}
\label{sec:results}

\subsection{Defect Reconstruction}

\begin{figure*}[!htb]
	\centering
    \includegraphics[scale=0.63]{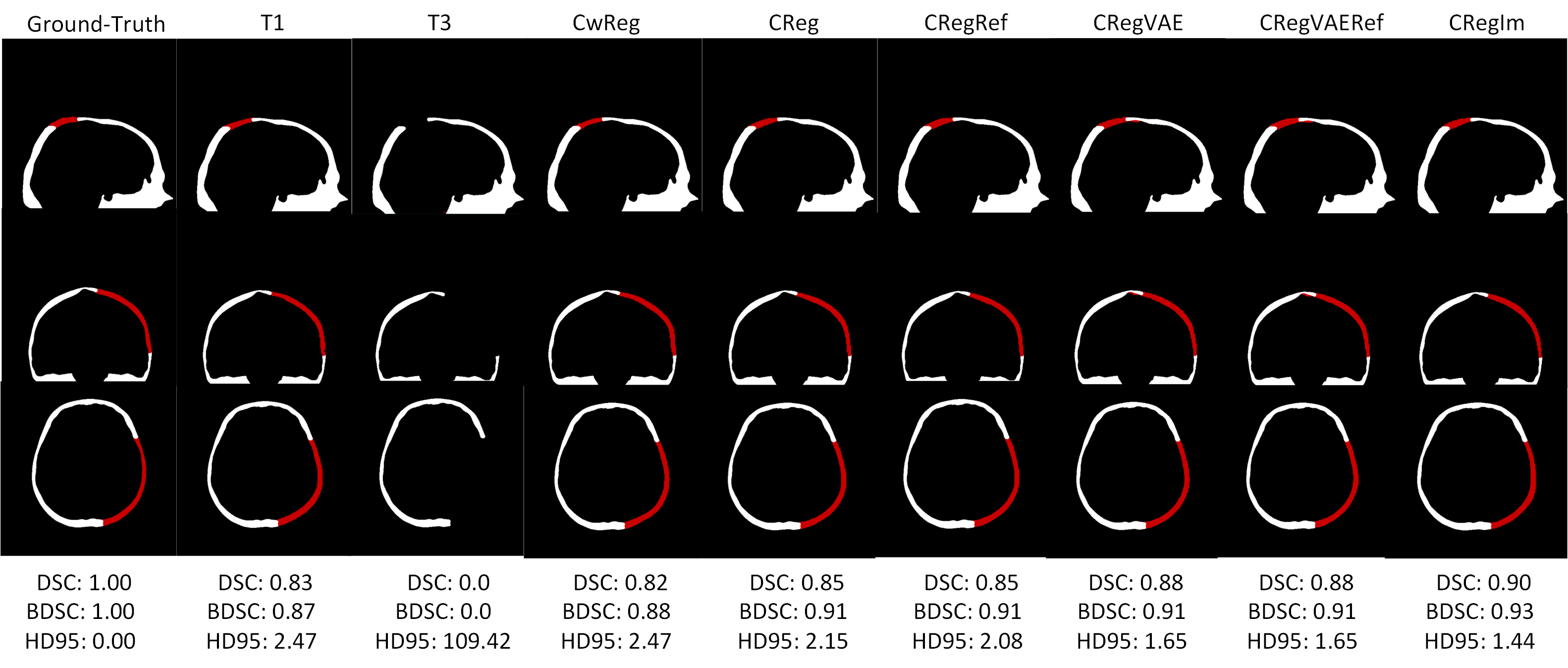}
    \caption{Exemplary results for case from Task 1 dataset - ID 15, Defect: Random\_2, (best viewed zoomed in electronic format).}  
    \label{fig:t1_vis}
\end{figure*}

\begin{figure}[!htb]
\centering
\begin{footnotesize}
	\begin{minipage}[b]{0.5\textwidth}
		\centering
		\centerline{\includegraphics[width=1\linewidth]{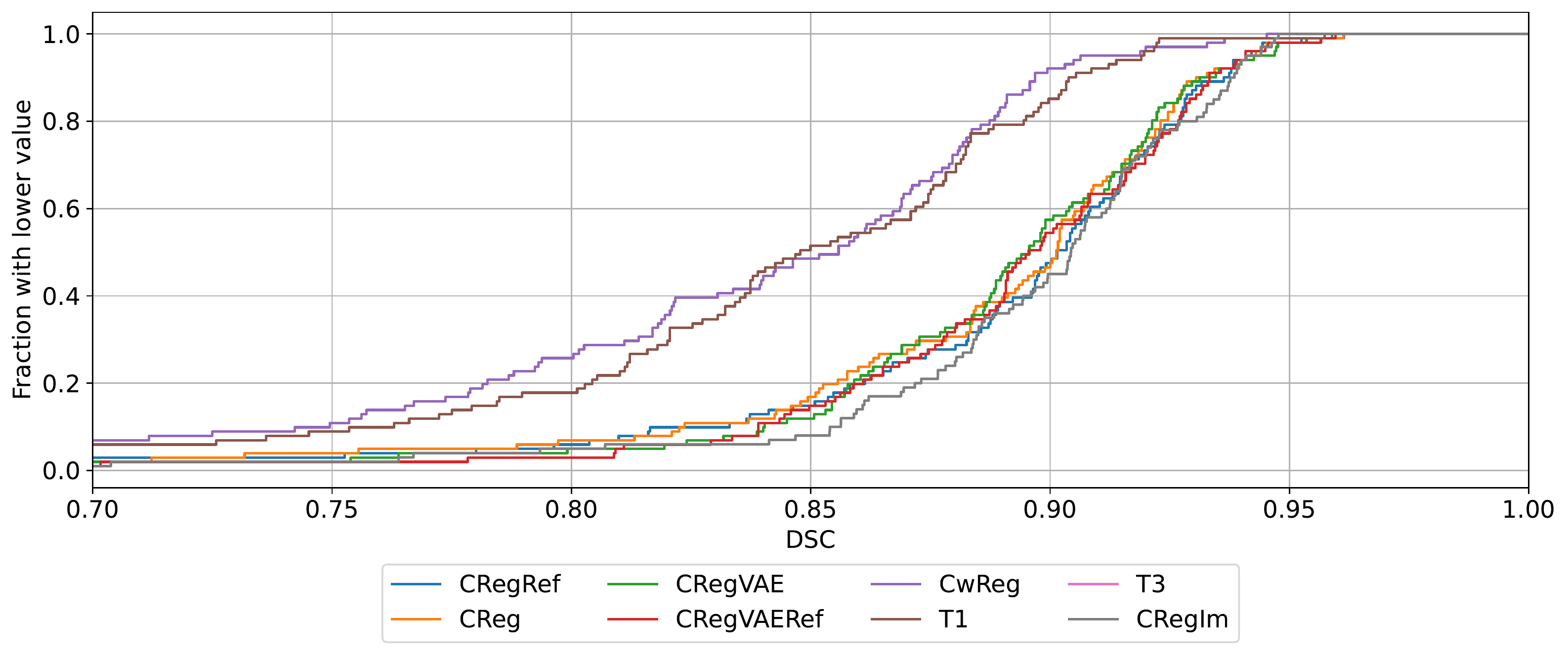}}
		\centerline{Task 1 - DSC}\medskip
	\end{minipage}
	\begin{minipage}[b]{0.5\textwidth}
		\centering
		\centerline{\includegraphics[width=1\linewidth]{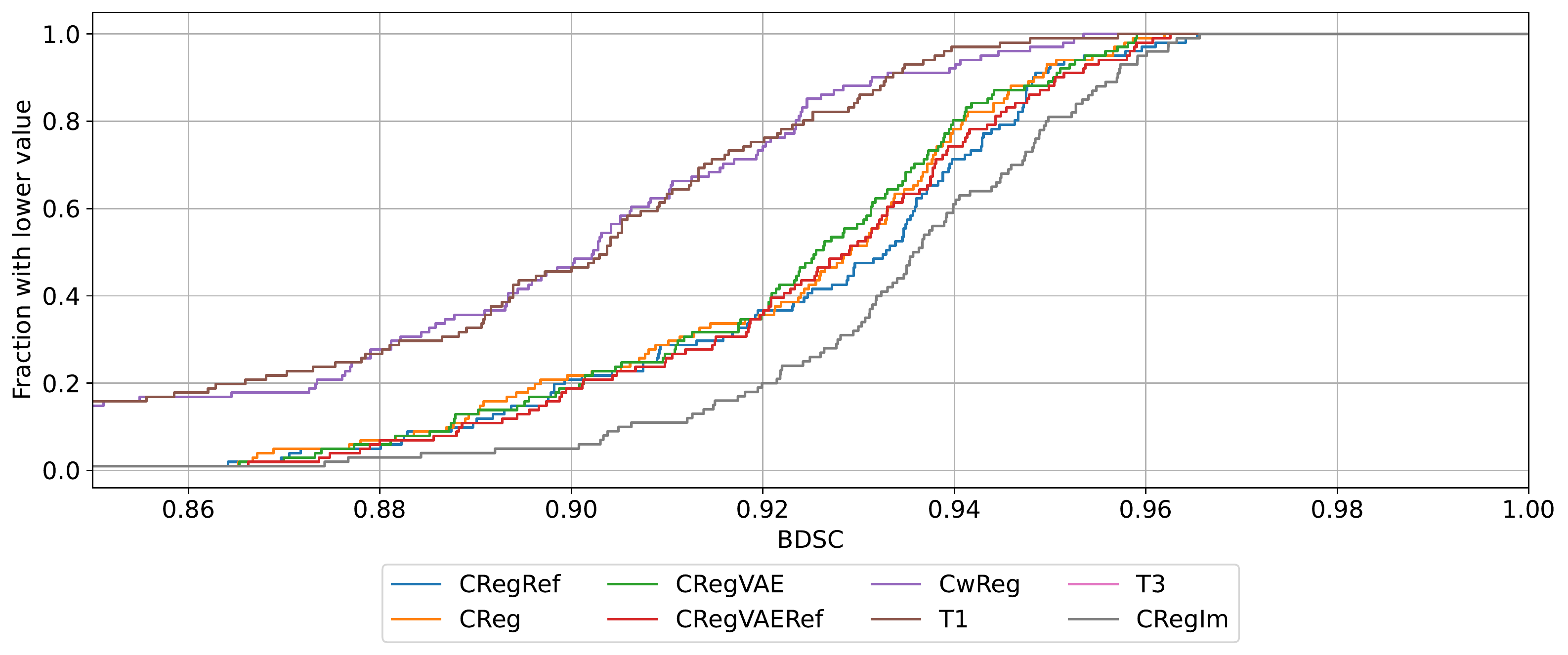}}
		\centerline{Task 1 - BDSC}\medskip
	\end{minipage}
		\begin{minipage}[b]{0.5\textwidth}
		\centering
		\centerline{\includegraphics[width=1\linewidth]{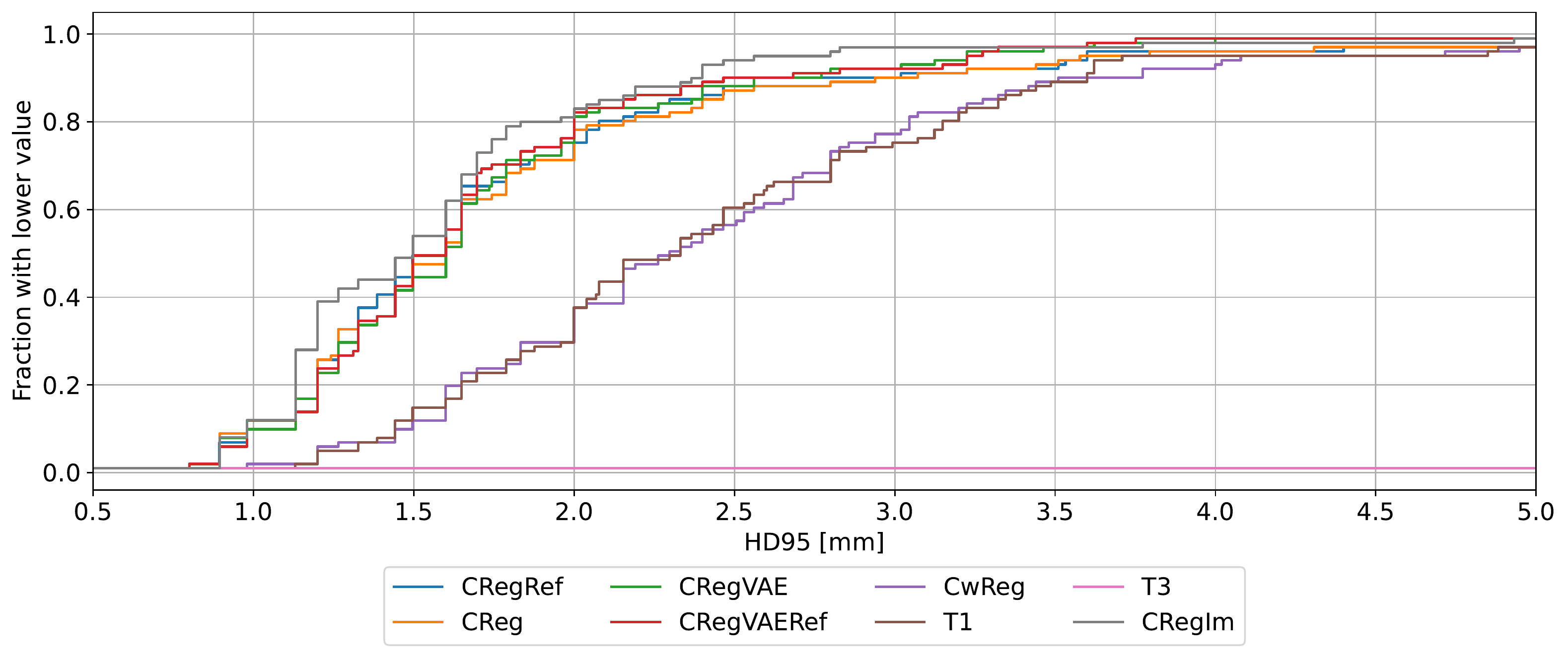}}
		\centerline{Task 1 - HD95 [mm]}\medskip
	\end{minipage}
\end{footnotesize}
\caption{Cumulative histograms for the Task 1 test set (presenting DSC, BDSC and HD95 respectively). Note the direct impact of imperfect registration on BDSC and HD95.}
\label{fig:t1_results}
\end{figure}

\begin{figure}[!htb]
\centering
\begin{footnotesize}
	\begin{minipage}[b]{0.5\textwidth}
		\centering
		\centerline{\includegraphics[width=1\linewidth]{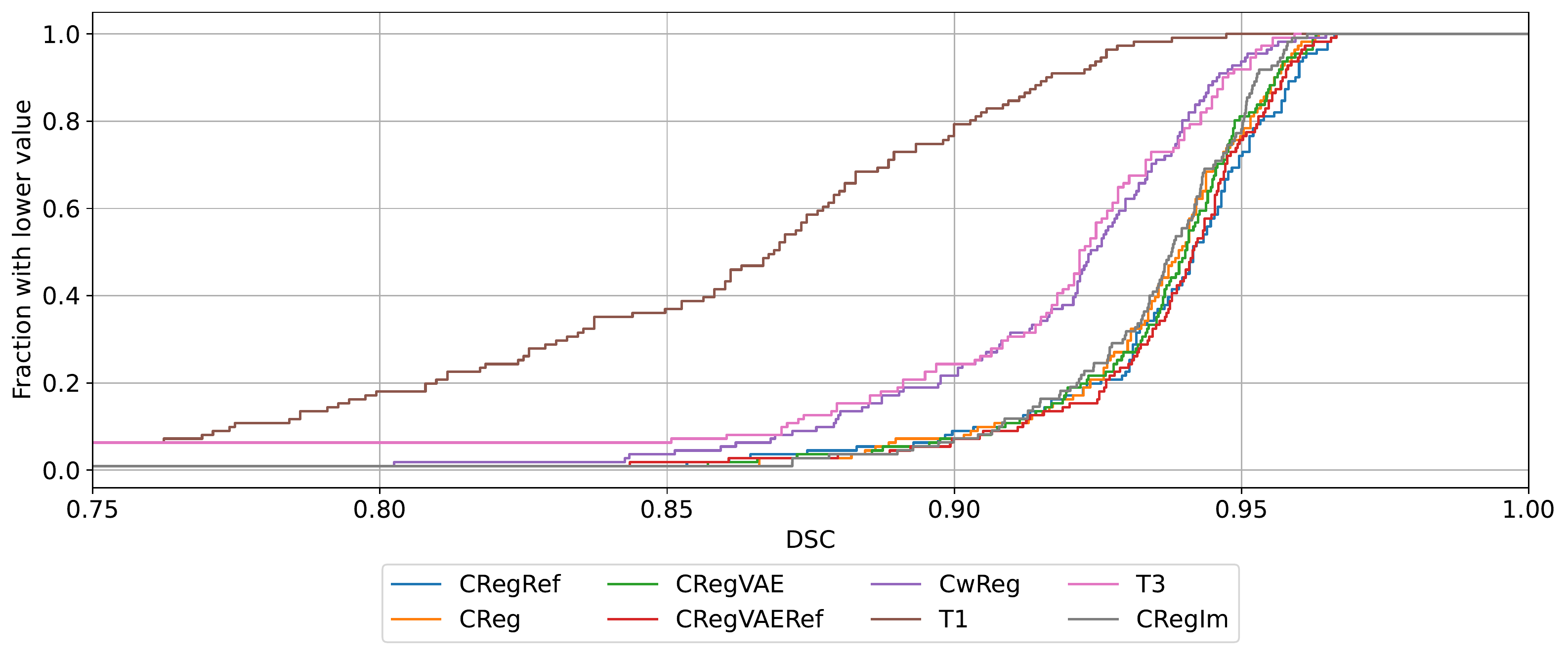}}
		\centerline{Task 3 - DSC}\medskip
	\end{minipage}
	\begin{minipage}[b]{0.5\textwidth}
		\centering
		\centerline{\includegraphics[width=1\linewidth]{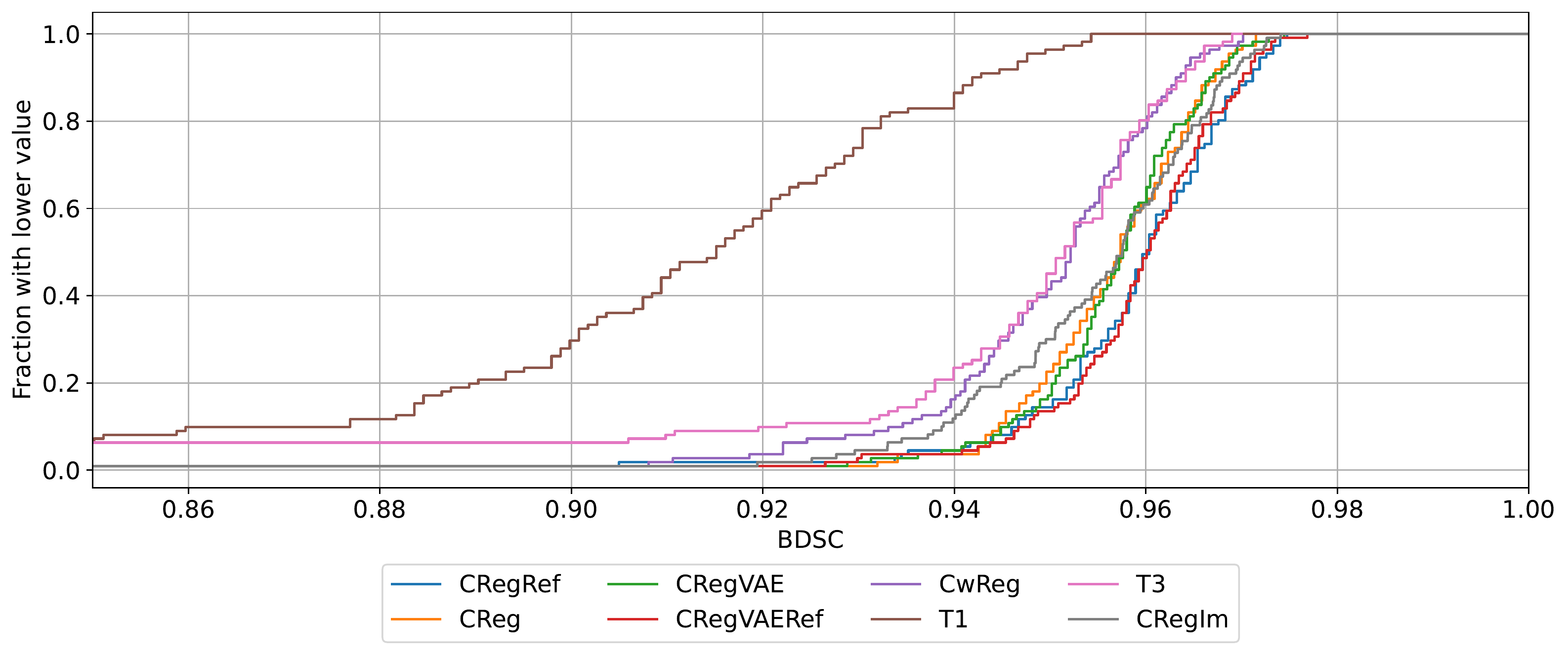}}
		\centerline{Task 3 - BDSC}\medskip
	\end{minipage}
		\begin{minipage}[b]{0.5\textwidth}
		\centering
		\centerline{\includegraphics[width=1\linewidth]{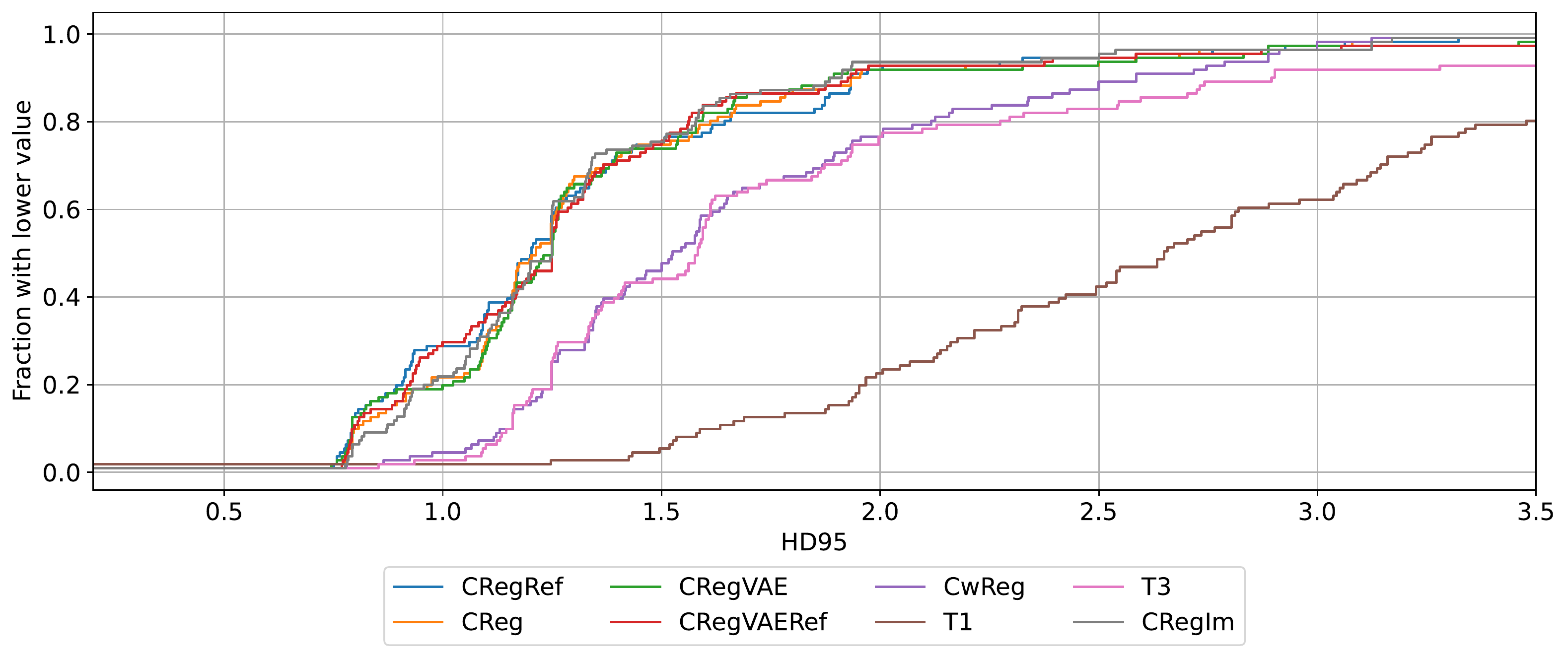}}
		\centerline{Task 3 - HD95 [mm]}\medskip
	\end{minipage}
\end{footnotesize}
\caption{Cumulative histograms for the Task 3 test set (presenting DSC, BDSC and HD95 respectively).}
\label{fig:t3_results}
\end{figure}

\begin{figure*}[!htb]
\centering
\begin{footnotesize}
	\begin{minipage}[b]{1.0\textwidth}
		\centering
		\centerline{\includegraphics[width=1\linewidth]{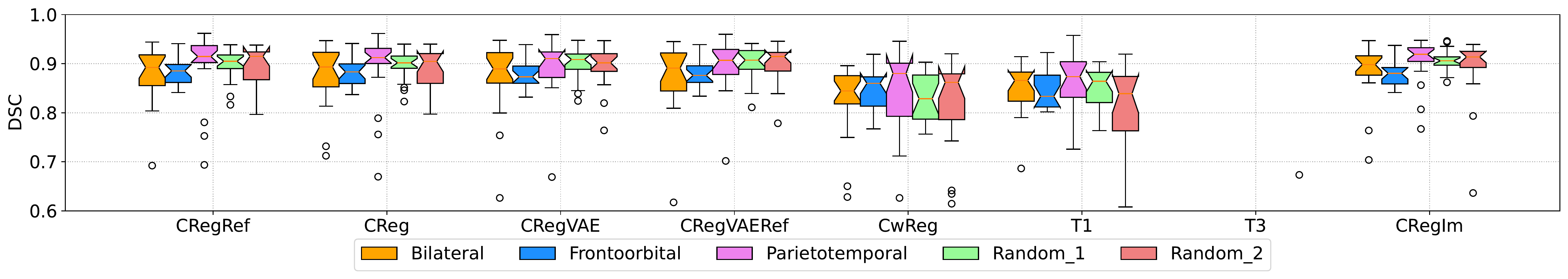}}
		\centerline{}\medskip
	\end{minipage}
	\begin{minipage}[b]{1.0\textwidth}
		\centering
		\centerline{\includegraphics[width=1\linewidth]{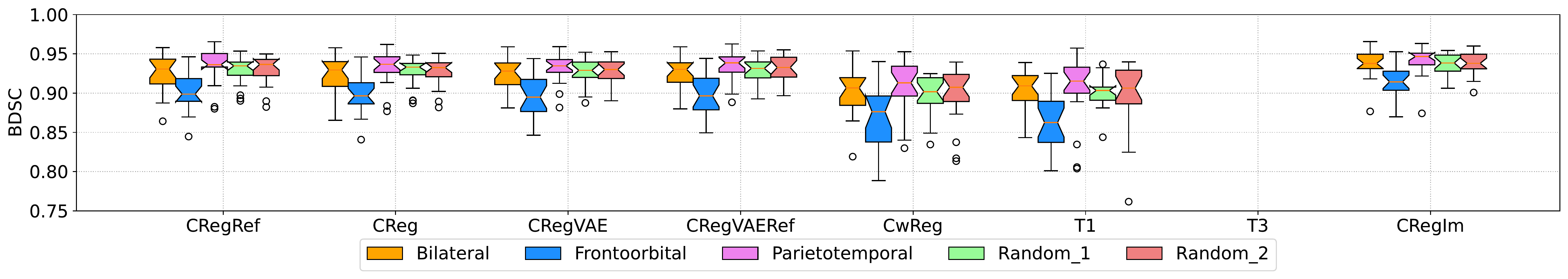}}
		\centerline{}\medskip
	\end{minipage}
	\begin{minipage}[b]{1.0\textwidth}
		\centering
		\centerline{\includegraphics[width=1\linewidth]{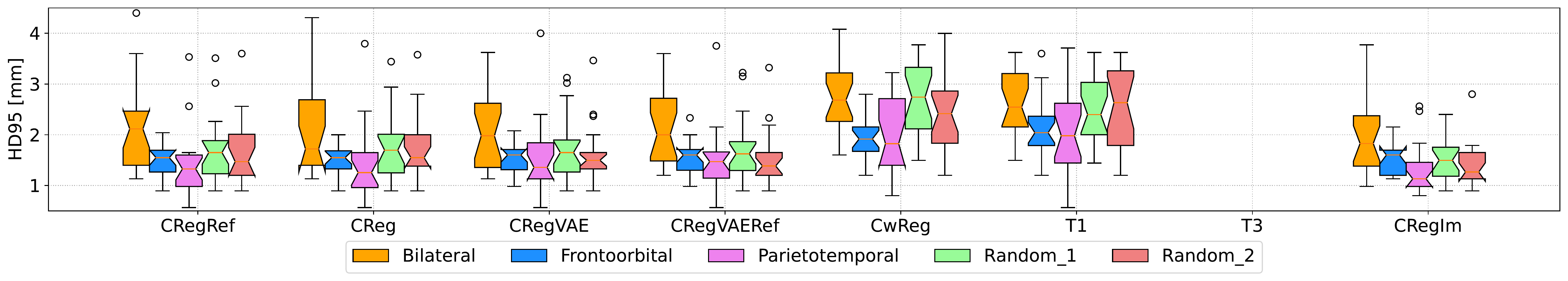}}
		\centerline{}\medskip
	\end{minipage}
\end{footnotesize}
\caption{Quantitative results for different defect types (Task 1). Note that the model trained using only Task 3 dataset is completely unable to generalize into Task 1 dataset.}
\label{fig:t1_bars}
\end{figure*}

The quantitative evaluation of the defect reconstruction is based on the Dice score (DSC), the boundary Dice score (BDSC), and the 95th percentile of Hausdorff distance (HD95). Figures~\ref{fig:t1_results},\ref{fig:t3_results} present cumulative histograms of the DC, BDSC, and HD95 for the Task 1 and Task 3 test sets separately. Figure~\ref{fig:t1_bars} presents the quantitative results for Task 1, including the division into different defect types. Table~\ref{tab:task_1_results} and Table~\ref{tab:task_3_results} show the quantitative statistics of different ablation studies, including other state-of-the-art methods. The exemplary visualization, presenting defective skulls together with the calculated defects, is shown in Figure~\ref{fig:t1_vis}.

\begin{table}[!htb]
\centering
\caption{Quantitative results for Task 1 test set presenting ablation studies and comparison to other state-of-the-art methods.}
\renewcommand{\arraystretch}{1.0}
\resizebox{0.7\textwidth}{!}{%
\begin{tabular}{cccc|cccc}
\label{tab:task_1_results}
Method & DSC & BDSC & HD95 [mm] & Method & DSC & BDSC & HD95 [mm] \\
\hline
\tabularnewline
T1 & 0.84 & 0.89 & 2.60 & T3 & 0.06 & 0.09 & 98.15 \\
CReg & 0.88 & 0.92 & 1.87 & CRegRef & 0.89 & 0.92 & 1.86 \\
CRegVAE & 0.89 & 0.92 & 1.74 & CRegVAERef & 0.89 & 0.0.93 & 1.71 \\
CwReg & 0.84 & 0.89 & 2.51 & CRegIm & \textbf{0.89} & \textbf{0.93} & \textbf{1.60} \\
\hline
Mahdi~\cite{mahdi_1} & 0.78 & 0.81 & 3.42 & Yang~\cite{yang_1} & 0.85 & 0.89 & 3.52 \\
\end{tabular}}
\end{table}

\begin{table}[!htb]
\centering
\caption{Quantitative results for Task 3 test set presenting ablation studies and comparison to other state-of-the-art methods. Please note that the method by Kroviakov~\textit{et al.}~\cite{kroviakov_1} was evaluated only on a subset of the test set. Note that the results for Task 3 are indistinguishable for majority of ablation studies since the synthetic defects are regular and similar.}
\renewcommand{\arraystretch}{1.0}
\resizebox{0.7\textwidth}{!}{%
\begin{tabular}{cccc|cccc}
\label{tab:task_3_results}
Method & DSC & BDSC & HD95 [mm] & Method & DSC & BDSC & HD95 [mm] \\
\hline
\tabularnewline
T1 & 0.81 & 0.86 & 10.59 & T3 & 0.87 & 0.91 & 4.40 \\
CReg & 0.93 & 0.95 & 1.96 & CRegRef & 0.93 & 0.95 & 1.93 \\
CRegVAE & 0.93 & 0.96 & 1.65 & CRegVAERef & 0.94 & \textbf{0.96} & 1.64 \\
CwReg & 0.92 & 0.95 & 2.00 & CRegIm & \textbf{0.93} & 0.95 & \textbf{1.47} \\
\hline
Mahdi~\cite{mahdi_1}  & 0.88 & 0.93 & 3.59 & Yu~\cite{yu_1} & 0.77 & 0.77 & 3.68 \\
Li~\cite{li_2} & 0.81 & - & - & Kroviakov~\cite{kroviakov_1} & 0.85 & 0.95 & 2.64 \\
Pathak~\cite{pathak_1} & 0.90 & 0.95 & 2.02 
\end{tabular}}
\end{table}

\subsection{Implant Modeling}

\begin{figure}[!htb]
	\centering
    \includegraphics[scale=0.67]{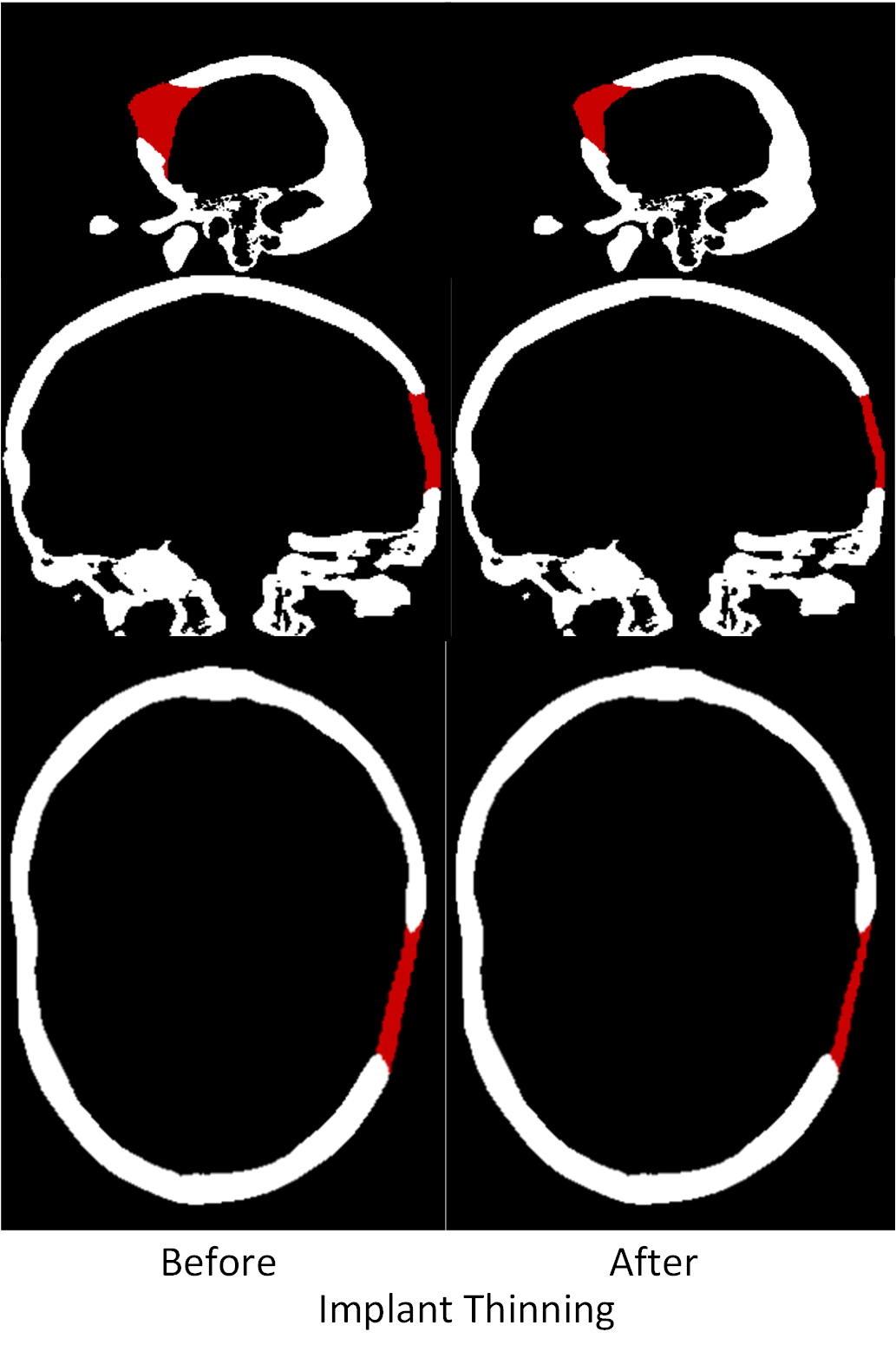}
    \caption{Exemplary visualization of the implant modeling (Task 2 - Case 10). For the quantitative results we refer to Table~\ref{tab:task_2_results}. Note that before the implant thinning the reconstruction is not implantable.}   
    \label{fig:thinning}
\end{figure}

The quantitative results of the implant modeling are presented in Table~\ref{tab:task_2_results}, for Task 2 only. There are no ground-truth implants for Task 1 and 3. It should be kept in mind that there may be multiple correct implants for a given defective skull (based e.g. on the desired implant thickness). Therefore the quantitative results are significantly lower than for the defect reconstruction. Additional expert assessment is required which was performed by the AutoImplant challenge organizers~\cite{autoimplant_2}. We show the effect of implant modeling in Figure~\ref{fig:thinning}.

\begin{table}[!htb]
\centering
\caption{Quantitative results for the Task 2 (real cranial defects), including implant modeling. Please note that the evaluation metrics are calculated with respect to the expert-designed implant, not the cranial defect. The implant modeling step improves results for all cases except the one for which the defect reconstruction step failed (ID 6).}
\renewcommand{\arraystretch}{1.0}
\resizebox{0.7\textwidth}{!}{%
\begin{tabular}{ccccccc}
\label{tab:task_2_results}
 & \multicolumn{3}{c}{Defect Reconstruction (CRegIm)} & \multicolumn{3}{c}{Implant Modeling}
\tabularnewline
Case & DSC & BDSC & HD95 [mm] & DSC & BDSC & HD95 [mm] \\
\hline
\tabularnewline
1 & 0.50 & 0.52 & 7.35 & 0.64 & 0.67 & 4.12 \\
2 & 0.60 & 0.57 & 7.48 & 0.75 & 0.71 & 3.60 \\
3 & 0.29 & 0.22 & 16.58 & 0.40 & 0.31 & 10.82 \\
4 & 0.51 & 0.48 & 10.72 & 0.66 & 0.63 & 5.48 \\
5 & 0.58 & 0.55 & 6.71 & 0.73 & 0.72 & 3.46 \\
6 & 0.53 & 0.66 & 14.73 & 0.50 & 0.68 & 15.07 \\
7 & 0.40 & 0.38 & 12.69 & 0.55 & 0.54 & 7.00 \\
8 & 0.49 & 0.41 & 15.56 & 0.63 & 0.54 & 9.00 \\
9 & 0.68 & 0.48 & 7.07 & 0.73 & 0.72 & 5.00 \\
10 & 0.59 & 0.58 & 6.16 & 0.73 & 0.66 & 3.00 \\
11 & 0.70 & 0.67 & 3.00 & 0.62 & 0.61 & 3.00 \\
\hline
Mean & 0.53 & 0.50 & 9.82 & \textbf{0.63} & \textbf{0.61} & \textbf{6.32} \\
Std & 0.12 & 0.13 & 4.48 & 0.11 & 0.12 & 3.86 \\
\hline
Yu \textit{et al.}~\cite{yu_1} & 0.52 & 0.45 & 8.34 & - & - & - \\
Mahdi \textit{et al.}~\cite{mahdi_1} & 0.38 & 0.33 & 51.24 & - & - & - \\

\end{tabular}}
\end{table}

\subsection{3-D Printing}

\begin{figure}[!htb]
	\centering
    \includegraphics[scale=0.7]{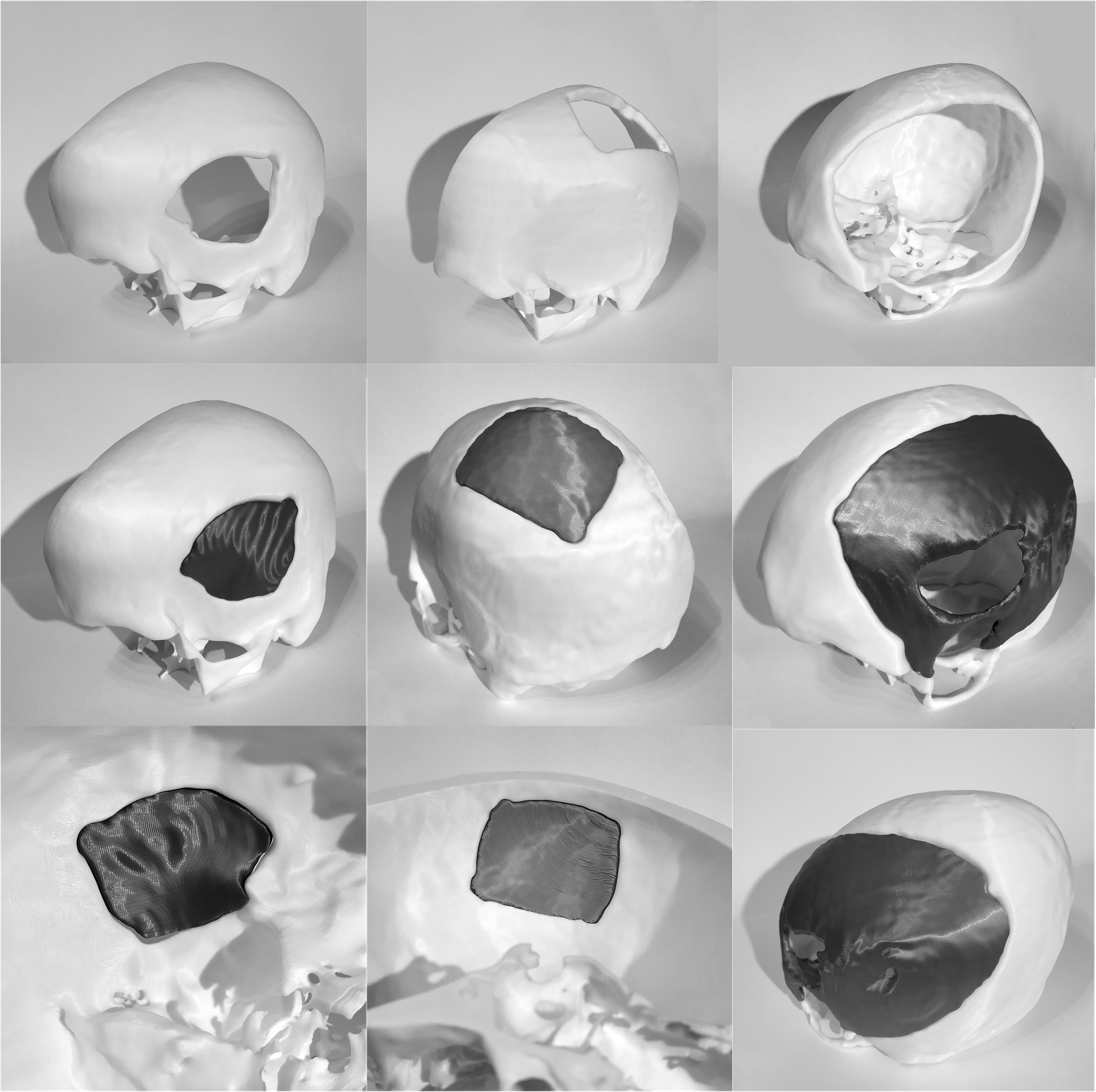}
    \caption{Exemplary 3-D printed skulls together with the automatically designed implants for real, cranial defects (Task 2). Note that one case (ID: 6) is modeled incorrectly. However, the case strongly differs from all other real/synthetic defects in terms of size and shape.}   
    \label{fig:3dprinting}
\end{figure}

Figure~\ref{fig:3dprinting} presents the skulls and the corresponding defects/implants for three cases from the Task 2 set. We decided to print three cases based on the qualitative results: the best, the worst, and the moderate one. Please note that for the moderate and the best case, the implants fill the cranial defect very well and the boundaries are smooth. The reconstruction worst case was unsuccessful due to the skull being significantly different than the training distributions and large defect size. Nevertheless, it may be observed that the implant is modeled correctly at the boundaries and requires just a minor manual modification to close the undesired hole.

\subsection{Mixed Reality}

\begin{figure}[!htb]
	\centering
    \includegraphics[scale=1.0]{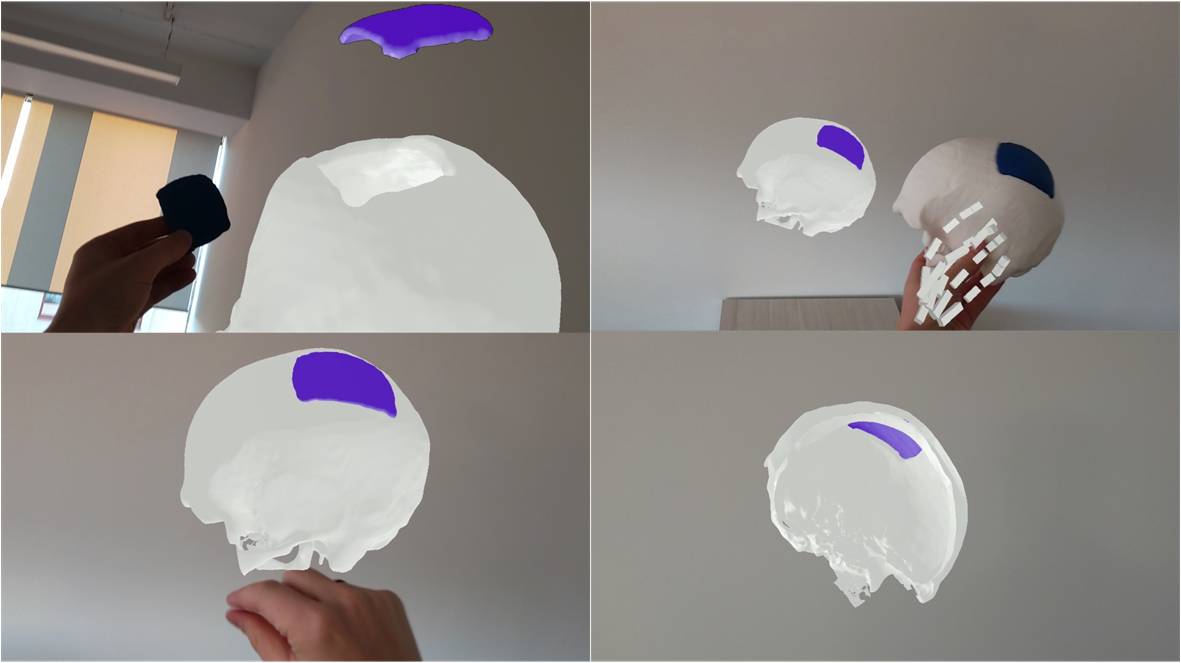}
    \caption{Exemplary visualization of the reconstructed defect in mixed reality (real cranial defect, Task 2). We refer to supplementary movie presenting the holographic interactions and system abilities.}   
    \label{fig:ar}
\end{figure}

Figure~\ref{fig:ar} shows a use case presenting the method outcome in mixed reality. The visualization is created using CarnaLife Holo software (MedApp S.A.). We show the implant hologram superimposed on the 3-D printed skull, and both the defective skull and the calculated implant as holograms. This kind of visualization is created almost instantly in contrast to the 3-D printed model. It enables the operating team to analyse the proposed implant structure and to plan the procedure in 3-D. Thus, it is possible to propose adjustments based on the medical experience of the operating team before the implant is printed. The intuitive real-time planning should result in reduction of possible errors. We also attach a supplementary movie presenting the procedure\footnote{\url{https://youtu.be/a1IMMtt3ovc}}.

\section{Discussion}
\label{sec:discussion}

\subsection{General Remarks}

The key component to improve the reconstruction accuracy and generalizability is connected with combining the datasets from different distributions. The best performing setup achieves DC, BDC, HD95 close to 0.91, 0.94, and 1.53 mm respectively. These results are capable of automating the cranial implant modeling and are the best among currently proposed methods. The pipeline allows one to create a personalized implant directly from the segmentation of the defective skull.

The quantitative results vary between the three test sets. The reason for this is connected with the heterogeneity of the dataset. Task 3 is rather straightforward since the defects are regular, with similar sizes and shapes (except the last 10 test cases). On the other hand, Task 1 introduces variability of the defect size and position. This, together with the fact that larger defects break the skull symmetry, leads to worse quantitative results. Interestingly, the variability between different defect types is minor and the method handles well all the defect variations. Only frontoorbital defects achieve considerably lower scores. It is related to breaking the skull symmetry. The quantitative scores for Task 2 are significantly different from the Task 1 and Task 3. However, they are not related to the reconstructions, but to the modeled implants. Since it is hard to define ground-truth implants, this evaluation requires the assessment of an expert neurosurgeon. 

\subsection{Ablation Studies}

The imperfect IR provides the most accurate results both from the quantitative and qualitative perspectives. The generalizability to real cranial defects is also the most promising. The augmentation by smooth and invertible IR or by the generative model improves the quantitative results on the Task 1/3 tests sets, however, generalize worse into data from different distributions. It is because the smooth and invertible IR changes only the defect shapes while maintaining the original skull structure. On the other hand, imperfect registration creates skulls in between the original pairs, effectively creating a more heterogeneous training set. We hypothesized that similar results could be achieved by VAE by lowering the influence of the KL divergence, however, it turned out that it increased the variety of generated skulls but the defects were mostly incomplete and not anatomically plausible. The same issue arose when the VAE was trained on the training set created by imperfect IR. The naive combination of training sets by a simple resampling to the same voxel size and resolution does not improve the skull reconstruction significantly. It can be also observed that an attempt to use just a single training set results in a lack of generalizability. This is well-shown in experiments using only Task 3 training data, which provide good results for the Task 3 test set, however, are completely unable to reconstruct defects from the remaining datasets.

Several reasons are motivating the aforementioned ablation studies. First, we decided to not include ablation studies concerning the network architecture and training hyperparameters. We observed that straightforward modifications to the network architecture (e.g. adding additional skip connections, increasing the number of layers or filters) do not influence the results significantly. The different setups of training hyperparameters influence only the training time. Eventually, the models converge to similar states. On the other hand, the crucial aspect to improve the results is connected with preparing and augmenting the training set. It shows that the proposed method's performance is limited mostly by the training set size and its homogeneity.

\subsection{Limitations}

The proposed method has several limitations. It can be observed that the quantitative results for the implant modeling are not as good as for the defect reconstruction. This is caused by the difficulty with defining the ground-truth implant, in contrast with the ground-truth for the shape completion, which is straightforward. The optimal shape for a given implant depends on external geometric properties and the mechanical properties of the 3-D printing material. Moreover, there may be several implants that may be successfully implanted. Therefore, the implant evaluation should be performed by an expert neurosurgeon, as in the AutoImplant summary publication~\cite{autoimplant_2}.

Another limitation is related to a significant difference between the DC and BDC for several cases. This happens when the reconstruction boundaries are well defined, however, the symmetry of the skull is lost. This is essential from an aesthetic point of view and could be addressed by further dataset extension or the use of statistical shape models~\cite{pimentel_1}.

In the current version, we assume that only a single implant is being modeled at a given time. However, this assumption is realized in postprocessing and may be easily relaxed to prepare several implants for a single skull with multiple defects.

\subsection{Significance and Practical Applications}

The automatic cranial implant design may improve several practical procedures. It may reduce the resources required for manual modeling. Only small adaptations may be required, thus lowering the time required for implant manufacturing. As a result, the automatic design enables 3-D printing directly during the craniectomy, resulting in a potential unnecessity to perform the follow-up intervention to fill the defect. The surgery procedures may be further improved by the use of mixed reality systems. We attach a supplementary video presenting the hologram of the exemplary implant together with a corresponding 3-D printed defective skull. This technology may aid surgeons during the craniectomy and potentially reduce the surgery time~\cite{cl_1}. Furthermore, we forecast that the presented pipeline and training strategy may be generalized into other areas, e.g. dental implants~\cite{bayakdar_1}.

\subsection{Future Work}


The most obvious way to proceed is to collect more clinical data samples and increase the training set variability. An example of such a dataset is MUG500+~\cite{MUG500}. We hope that in the next edition of the AutoImplant challenge, the organizers will preprocess the dataset by creating synthetic defects, and a unified representation of the ground-truth implants. Currently, the ground-truth implants are released in STL format. To make the research more reproducible, the ground-truth implants should be released in binary format, similarly to the defective skulls. However, as discussed in the Limitations section, it is hard to define which implant is the most suitable and it cannot be done at the level of the binary mask alone. Furthermore, the evaluation should be performed on cranial defects from different ethnic groups to ensure fairness and inclusivity. 

The training set augmentation could be further explored. It would be interesting to compare several generative models to observe the influence on the results within the same and different distributions. In this research, we used VAEs due to the well-defined problem and their training stability. However, the generative adversarial networks (GANs) or GAN-VAEs could further increase the training set heterogeneity. Moreover, it would be beneficial to quantitatively describe the influence of the IR quality on the defect reconstruction. However, this is computationally intensive since the IR-based augmentation requires thousands of 3-D registrations and increases quadratically with the number of skulls.

The use of deep reinforcement learning could be also interesting. This approach may limit the necessity of the shape completion step. The implant could be modeled directly using a set of pre-defined rules. These rules may be connected with both the geometric properties (e.g. minimum and maximum implant thickness, possibility to be implanted) and the mechanical properties of a given material. However, this approach requires significant computational resources during the training phase.

Another research area, strongly connected with the problem discussed, is finding the optimal scheme for the implants postprocessing to prepare them for 3-D printing. It would be beneficial to propose a pipeline to completely automate the printing procedure.

\section{Conclusion}
\label{sec:conclusion}

In this work, we proposed a complete pipeline to perform the automatic cranial defect reconstruction and implant modeling. We evaluated the proposed method on public datasets and obtained results significantly higher than the state-of-the-art methods. We released the source code to support the experiment's reproducibility. We performed several ablation studies, presented the method limitations, discussed practical use cases, and defined future research directions. The presented method is a significant contribution to the automatic design of cranial implants and may be potentially adapted to other implant types.


\section*{Acknowledgment and Compliance With Ethical Standards}

This research study was conducted retrospectively using human subject data made available in open access by AutoImplant 2021 challenge organizers~\cite{dataset}. The authors declare no conflict of interest.

We thank the AutoImplant organizers for evaluating our ablation studies on the closed test set, and the MedApp S.A for giving us an access to the CarnaLife Holo software to create the visualizations in mixed reality.

\section*{Author Contributions}

M.W (65\%): Conceptualization of this study, Methodology, Software, Experiments, Article preparation, Evaluation, Visualizations, Analysis of the results, Critical revision.
M.D (10\%): High performance computing, Article preparation, Evaluation, Visualizations, Critical revision.
M.So (10\%): 3-D printing, Article preparation, Critical revision.
D.H (5\%): Article preparation, Supervision, Critical revision.
M.St (5\%): Mixed reality use case, Critical revision.
A.S (5\%): Mixed reality use case, Article preparation, Supervision, Critical revision.

\FloatBarrier

\bibliographystyle{ieeetr}
\bibliography{main}

\end{document}